\documentclass[aps,prb,superscriptaddress,twocolumn,floatfix]{revtex4}
\usepackage{graphicx}
\usepackage{amsmath}
\usepackage{amssymb}
\def\opex{ Opt.\ Express }

\begin{document}

\title{Enhanced stimulated Raman scattering in slow-light photonic crystal waveguides}
\author{James F. McMillan, Xiaodong Yang, Nicolae C. Panoiu, Richard M. Osgood, and Chee Wei Wong}
\address{Optical Nanostructures Laboratory, Columbia University, New York, NY 10027}

\begin{abstract}
We investigate for the first time the enhancement of the stimulated Raman scattering in slow-light
Silicon-on-Insulator (SOI) photonic crystal line defect waveguides. By applying the Bloch-Floquet
formalism to the guided modes in a planar photonic crystal, we develop a formalism that relates
the intensity of the down-shifted Stokes signal to the pump intensity and the modal group
velocities. The formalism is then applied to two prospective schemes for enhanced stimulated Raman
generation in slow-light photonic crystal waveguides. The results demonstrate a maximum factor of $10^{4}$
(66,000) enhancement with respect to SOI channel waveguides. Effects of two photon absorption,
intrinsic scattering, and disorder with respect to slow-light Raman generation towards
optically-pumped silicon amplifiers and lasers are also discussed.

\end{abstract}

{\footnotesize\sf
Submitted to {\it Optics Letters} (November, 2005) }
\maketitle

Silicon photonics has seen remarkable advancements in recent years. Subwavelength silicon
nanostructures - such as photonic crystals and high-index-contrast photonic integrated circuits -
offer the opportunity to manipulate the propagation of light at sub-wavelength scales. Moreover, the
inherent ease of integrating the silicon photonics platform with CMOS foundry ICs offers unprecedented bandwidth 
per unitcost and distance in optical data communications.

\begin{figure}[b]
\centering \leavevmode
\includegraphics[]{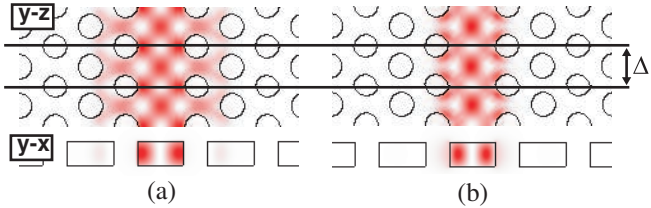}
\caption{\label{fmodes}Calculated bound states of a hexagonal lattice W1 PhCWG with defect modes
separated by the LO/TO optical phonon (Scheme 1). a)Stokes b)Pump}
\end{figure}

Silicon, however, is at an intrinsic disadvantage for optical amplification and lasing due to its
indirect band gap and non-existent second order nonlinear response. Recent work has demonstrated
that stimulated Raman scattering (SRS) in single-crystal silicon channel waveguides is a feasible
means to achieve amplification and lasing \textit{via} optical pumping
\cite{edo04oe,lia04apl,xu04opex,bj04oe,rlj05n}. This is due to the intrinsically large Raman
gain coefficient in silicon (being $10^{3}$ to $10^{4}$ times greater than for silica) and
silicon nanostructures offering the benefit of high optical confinement due to high-index contrast of silicon
with air or silicon oxide. While still requiring an optical pump and possessing limited gain
bandwidth, enhanced SRS through slow-light silicon photonic crystal waveguides (PhCWG) can serve
as an ultra-compact on-chip gain media at desired telecommunications frequencies. Enhanced Raman
scattering has been observed in bulk hollow-core slow-light guided-wave structures \cite{kan02jetpl}
and has also recently been suggested for photonic crystal (PhC) defect nanocavities.\cite{yw05oe}
In addition, a semiclassical model of Raman scattering in bulk photonic crystals has been
introduced.\cite{fz05pre} In this Letter we demonstrate theoretically for the first time the
explicit enhancement of SRS in a slow-light PhCWG through a four wave mixing formalism from the
computed modes of the line-defect waveguide.
\begin{figure} [b]
\centering \leavevmode
\includegraphics[]{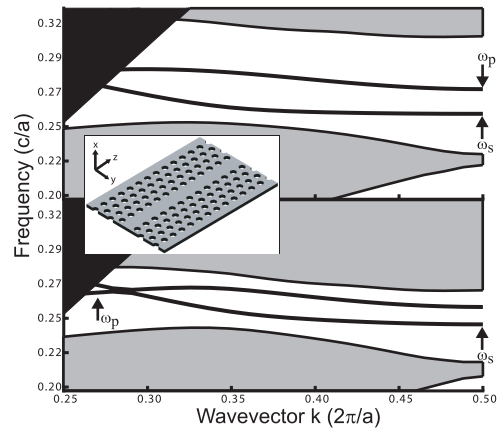}
\caption{\label{fbs}Projected band structure of silicon W1 PhCWG indicating pump and Stokes
frequencies. (top) Scheme 1, ($r/a=0.29$) (bottom) Scheme 2,  ($r/a=0.22$). ($h/a=0.6$) in both
cases. (inset) W1 PhCWG.}
\end{figure}
A silicon PhCWG studied here, made by removing a single row in a hexagonal lattice of holes -
denoted as "W1 PhCWG" - and its projected band structure can be seen in Fig. \ref{fbs}. This
structure supports two tightly confined modes with small group velocities, as illustrated by the
two bands within the band gap, with frequencies below the light line. The field distribution of
these two modes, as computed through the plane wave expansion method,\cite{jj01oe} is illustrated
in Fig. \ref{fmodes}. The strong subwavelength modal confinement of the high index contrast PhCWG
leads to increased field intensities in the silicon gain media, permitting increased nonlinear
interactions. In addition to increased field intensities from high index confinement, there is
additional SRS enhancement from the small group velocities of the PhCWG propagating modes.
Physically this enhancement originates from the effective long light-matter interaction times at small group
velocities. Photon localization is observed at the band
edge; the photon experiences multiple scattering processes and moves very slowly
through the material structure. The guided bands of a 2D PhCWG can be designed
to be as flat as desired ($v_{g}\equiv d\omega/dk$) for slow-light behavior, and group velocities
as low as $10^{-2}c$ to $10^{-3}c$ have been demonstrated.\cite{nys01prl,gke05prl}

In SRS for silicon an incident photon interacts with the LO and TO phonons. The strongest
Stokes peak arises from the single first-order Raman-phonon at the center of the Brillouin zone.
The generation of the Stokes photons can be understood classically as a third order nonlinear
effect and this formalism has been used to model SRS in SOI waveguides, both in CW \cite{dhc03ol}
and pulsed\cite{cpo06jqe} operation. It can be modeled in bulk materials as a degenerate four-wave-mixing
problem involving the pump and Stokes beams. The important material parameter is the third
order nonlinear Raman susceptibility, $\chi^{R}$. For silicon, at resonance, $\chi^{R}$ is defined
by the components $\chi^{R}_{ijij}$ = -$i\chi^{R}$ = -$i11.2\times10^{-18}~\mathrm{m}^{2}\cdot
\mathrm{V}^{-2}$ ($i,j=1,2,3$). An additional symmetry, imposed by the crystal point group
(\textit{m3m} for Si), is $\chi^{R}_{iijj}$ = 0.5$\chi^{R}_{ijij}$. These components, and their
permutations as defined by the crystal point group, define the SRS in a silicon crystal.
For our purpose we shall consider scattering in silicon along the
[$1\bar{1}0$] direction since practical devices are fabricated along this direction due to the
favorable cleaving of silicon along this direction.

For bulk silicon, the evolution of the Stokes beam is defined by the following equation \cite{agr89ac}
\begin{equation}
\label{bulkeq} \frac{\displaystyle dI_{s}}{\displaystyle d z} = -\frac{\displaystyle
3\omega_{s}\mathrm{Im}(\chi^{R}_{\mathrm{eff}})}{\displaystyle
\epsilon_{0}c^{2}n_{p}n_{s}}I_{p}I_{s},
\end{equation}
where $\chi^{R}_{\mathrm{eff}}=\sum_{ijkl}\chi^{R}_{ijkl} \mathbf{\hat{\alpha}}^{*}_{i}
\mathbf{\hat{\beta}}_{j} \mathbf{\hat{\beta}}_{k} \mathbf{\hat{\alpha}}_{l}$. Here,
$\mathbf{\hat{\alpha}}$ and $\mathbf{\hat{\beta}}$ are unit vectors along the polarization
directions of the pump and Stokes beams, respectively. Eq. (\ref{bulkeq}) describes the gain of
the Stokes intensity, $I_{s}$.  It shows an intrinsic dependence on the polarization and the
phonon selection rules through $\chi^{R}$, and the intensity of the pump beam by $I_{p}$.  The
bulk solution also describes SRS in dielectric waveguides, where $\chi^{R}_{\mathrm{eff}}$ is
averaged over the waveguide mode field distribution.

A PhCWG presents a very different field distribution than the bulk or dielectric waveguide case.
As shown in the computed modal profiles of Fig. \ref{fmodes}, the mode differs from that of a
conventional channel waveguide in that it exhibits a periodic variation in the direction of
propagation. We introduce the modal distribution of the pump and Stokes modes in a Bloch-Floquet
formalism, \label{modes}
\begin{equation}
\mathbf{\mathcal{E}}_{n,\mathbf{k}_{n}}(\mathbf{r},\omega_{n}) =
\mathbf{E}_{n,\mathbf{k}_{n}}(\mathbf{r},\omega_{n})
\exp[i\mathbf{k}(\omega_{n})\cdot\mathbf{r}],\label{modes}
\end{equation}
where $n$ is a mode index $(n=p,s)$, $\mathbf{k}_{n}=\mathbf{k}(\omega_{n})$ is the mode
wave-vector, $\mathbf{E}_{n,\mathbf{k}_{n}}(\mathbf{r},\omega_{n})$ is the modal distribution
within a unit cell of the PhC, defined in Fig. \ref{fmodes}, and obeys Bloch boundary condition
$\mathbf{E}_{n,\mathbf{k}_{n}}(\mathbf{r}+\mathbf{\Delta},\omega_{n}) =
\mathbf{E}_{n,\mathbf{k}_{n}}(\mathbf{r},\omega_{n})$. $\Delta$ defines the length of the unit
cell in the direction of propagation and for a W1 waveguide this equals the PhC lattice constant
$a$. To develop an equation that relates the evolution of the Stokes mode to the pump mode, we
employ the Lorentz reciprocity theorem,\cite{booksl}
\begin{equation}\label{Lorentz}
\frac{\displaystyle \partial}{\displaystyle \partial z} \int_{A} \big [
\mathbf{E}_{n,\mathbf{k_n}}^{*} \times \mathbf{\tilde{H}} + \mathbf{\tilde{E}} \times
\mathbf{H}_{n,\mathbf{k_n}}^{*}\big ] \cdot \mathbf{\hat{e}}_{z} dA = i\omega\int_{A} \mathbf{P}^{R}
\cdot \mathbf{E}_{n,\mathbf{k_n}} dA,
\end{equation}
This relates the unperturbed PhCWG modes of the pump or Stokes wavelengths,
$\{\mathbf{E}_{n,\mathbf{k}_{n}},\mathbf{H}_{n,\mathbf{k}_{n}}\}$, to those of the nonlinearly
induced fields. The envelopes of the fields are defined as
\begin{subequations}
\label{fields}
\begin{eqnarray}
\label{fieldE}&\mathbf{\tilde{E}}(\mathbf{r}) =
u_{s}(z)\mathbf{E}_{s,\mathbf{k}_{s}}(\mathbf{r},\omega_{s}) +
u_{p}(z)\mathbf{E}_{p,\mathbf{k}_{p}}(\mathbf{r},\omega_{p}), \\
\label{fieldH}&\mathbf{\tilde{H}}(\mathbf{r}) =
u_{s}(z)\mathbf{H}_{s,\mathbf{k}_{s}}(\mathbf{r},\omega_{s}) +
u_{p}(z)\mathbf{H}_{p,\mathbf{k}_{p}}(\mathbf{r},\omega_{p}),
\end{eqnarray}
\end{subequations}
with the assumption that the change in the pump and Stokes field amplitudes, $u_{p}(z)$ and
$u_{s}(z)$ respectively, over the length of the unit cell of the waveguide is very small
($\Delta\frac{ d u_{p,s}}{d z}\ll 1$). Taking the fields as defined in Eq.
(\ref{fields}), we derive the dependence of the Stokes amplitude on the longitudinal distance,
$z$,
\begin{equation}\label{zdep}
\frac{\displaystyle d u_{s}(z)}{\displaystyle d z} = \frac{\displaystyle
i\omega_{s}}{\displaystyle 4P_{s}\Delta} \int_{V_{0}} \mathbf{P}^{R}(\mathbf{r},\omega_{s}) \cdot
\mathbf{E}_{s,\mathbf{k}_{s}}(\mathbf{r},\omega_{s}) dV,
\end{equation}
where $P_s$ is the mode power and $\mathbf{P}^{R}(\mathbf{r},\omega_{s}) = 6\epsilon_{0} \hat{\chi}^{R} \vdots
\mathbf{E}^{*}_{p,\mathbf{k}_{p}}(\mathbf{r}) \mathbf{E}_{p,\mathbf{k}_{p}}(\mathbf{r})
\mathbf{E}_{s,\mathbf{k}_{s}}(\mathbf{r}) |u_{p}|^{2}u_{s}$. The integral in Eq. (\ref{zdep}) is
taken over the volume ($V_{0}$) of the unit cell of the PhCWG mode. Furthermore, the group
velocity of the modes can expressed by the following equation \cite{booksl}
\begin{equation}\label{groupv}
v_{g}^{p,s} = \frac{\displaystyle P_{p,s}\Delta}{\displaystyle \frac{\displaystyle
1}{\displaystyle 2}\epsilon_{0}\int_{V_{0}}\epsilon(\mathbf{r})
|\mathbf{E}_{p,s}(\mathbf{r},\omega_{p,s})|^{2}dV},
\end{equation}
With Eqs. (\ref{fields}) and (\ref{groupv}), and by rewriting Eq. (\ref{zdep}) in terms of the
modes intensity, an equation for the intensity of the Stokes mode inside the
PhCWG is obtained,
\begin{equation}
\label{PCeqI} \frac{\displaystyle dI_{s}}{\displaystyle d z} = -\frac{\displaystyle
3\omega_{s}}{\displaystyle \epsilon_{0}v_{g}^{p}v_{g}^{s}}\kappa I_{p}I_{s}.
\end{equation}
where
\begin{equation}
\label{kappa}\kappa = \frac{\displaystyle \Delta A_{\mathrm{eff}}\mathrm{Im}\bigg ( \int_{V_{0}}
\mathbf{E}^{*}(\omega_{s}) \cdot \hat{\chi}^{R} \vdots \mathbf{E}^{*}(\omega_{p})
\mathbf{E}(\omega_{p}) \mathbf{E}(\omega_{s}) dV \bigg )}{\displaystyle \bigg (
\frac{\displaystyle 1}{\displaystyle 2}\int_{V_{0}}\epsilon(\mathbf{r})
|\mathbf{E}(\omega_{p})|^{2}dV \bigg )\bigg ( \frac{\displaystyle 1}{\displaystyle
2}\int_{V_{0}}\epsilon(\mathbf{r}) |\mathbf{E}(\omega_{s})|^{2}dV \bigg )}
\end{equation}
is the efffective susceptibility. Here, the effective area $A_{\mathrm{eff}}$ is defined as the average modal area across the volume
$V_{0}$,
\begin{equation}
\label{Aeff}A^{2}_{\mathrm{eff}} = \frac{\displaystyle \bigg ( \int_{V_{0}}x^{2}
|\mathbf{E}(\omega_{s})|^{2}dV \bigg )\bigg ( \int_{V_{0}}y^{2} |\mathbf{E}(\omega_{s})|^{2}dV
\bigg )}{\displaystyle \bigg ( \int_{V_{0}} |\mathbf{E}(\omega_{s})|^{2}dV {\bigg )}^{2}}.
\end{equation}
The final equation, Eq. (\ref{PCeqI}), shows the explicit inverse dependence the Stokes mode
amplification has on the group velocities of the pump and Stokes modes.

Table \ref{coeff} shows the results of Eq. (\ref{PCeqI}) as being applied to two different PhCWG
schemes for SRS. The group velocities are calculated from the slope of the projected band
structure. The first (Scheme 1) involves utilizing both the guided modes of the W1 waveguide;
odd-parity is the pump mode and even-parity is the Stokes mode. The wavelength separation of the
modes at the edge of the Brillioun zone is matched to the LO/TO frequency separation of the pump
and Stokes beams (15.6 THz in Si \cite{th73prb}). The second (Scheme 2) utilizes a wide bandwidth
PhCWG,\cite{dmv05prb} in order for the Stokes and pump modes to exist both in the fundamental mode
and below the light line. The arrows in Fig. \ref{fmodes} indicate the pump and Stokes frequency
locations for both schemes.
\begin{table}[b]
  \centering
  \caption{\textbf{Group velocity and Effective Susceptibility in PhCWG schemes}}\label{coeff}
\begin{tabular}{|c|c|c|c|}
  \hline
  Scheme & $v_{g}^{s}$ & $v_{g}^{p}$ & $\kappa (\times 10^{-19}) [\mathrm{m}^{2}\cdot
\mathrm{V}^{-2}]$ \\
  \hline
  1 & 0.00017c & 0.0077c & 0.55 \\
  2 & 0.0041c & 0.24c & 2.02 \\
  \hline
\end{tabular}
\end{table}

From the results of Table \ref{coeff}, the Raman gain - proportional to
$\kappa/v_{g}^{p}v_{g}^{s}$ - is enhanced by up to approximately $10^{4}$ (Scheme 1:66,000, Scheme 2:86) times compared to
bulk Si based on a comparison of the respective group velocities. The results in Table \ref{coeff}
also show a $\kappa$ value of the same order with a conventional SOI waveguide.\cite{cpo06jqe} In
addition, we note a reduction in $\kappa$ in Scheme 1 as compared to Scheme 2, due to the lower
modal overlap. However, the single mode (Scheme 2) operation has the disadvantage that only the
Stokes mode, and not both modes, are at low group velocities for enhanced SRS.

The above results highlight the benefits of SRS enhancement through slow-light interactions in
compact PhCWG schemes. This approach can be readily extended to include two photon and bulk free
carrier absorption effects \cite{cpo06jqe} which may limit the effective Raman gain in PhCWGs.
These effects, in the experimental realization of silicon SRS amplification and lasing in
slow-light PhCWGs, can be surmounted with pulsed-laser operation \cite{bj04oe} or PIN diodes
\cite{rlj05n} to sweep the free-carriers.

In addition, we note recent theoretical\cite{hry05prl} and
experimental\cite{vm06ol} studies of PhCWGs, which show that slow group
velocity modes exhibit increased scattering losses. These losses are from coupling and intrinsic
(backscatter) reflection. Coupling into slow-light modes is currently the dominant loss
experimentally, although this can in principle be reduced through careful adiabatic coupling
\cite{jbs02pre} between the PhCWGs and input/output channel bus waveguides. Moreover, with thorough attention
to fabrication disorder, reflection
losses in PhCWG are suggested to be comparable with index-guided waveguides\cite{mlpov04apl}.These
scattering losses can thus potentially be smaller than the enhanced SRS gain discussed, permitting
the possibility for compact silicon Raman amplifiers and lasers. We also note that, for the same
desired Raman gain, the device length is significantly reduced, by $(c/v_{g})^{2}$, allowing
compact integration for high-density photonic circuits.\\

This research was sponsored in part by DARPA, and the Columbia
Initiatives in Science and Engineering in Nanophotonics. Both NCP and RMO would like to
acknowledge financial support through AFOSR through contracts FA95500510428 and FA9550-04-C-0022,
and NSF, Grant no. ECS-0523386. The authors also thank Steven G. Johnson for useful discussions on
the low group velocity scattering. C.~W. Wong's email address is cww2104@columbia.edu.

\end{document}